\documentclass[preprint]{emulateapj}

\usepackage{epsfig}
\usepackage{graphics}
\usepackage{float}
\usepackage[figuresright]{rotating}
\usepackage{amsmath}
\usepackage{multirow}
\usepackage{longtable}
\usepackage{rotate}
\usepackage{amssymb}
\usepackage{array}
\def\mytablecomments#1{\par\smallskip\noindent Notes. #1}

\shorttitle{SN~2004dj}
\shortauthors{Kotak et al.}

\def\kms{\ifmmode{\rm km\,s^{-1}}\else\hbox{$\rm km\,s^{-1}$}\fi}

\begin{document}


\title{Early-time {\it Spitzer\/} observations of the Type II-Plateau supernova, 2004dj}


\author{Rubina Kotak\altaffilmark{1}, Peter Meikle,  }
\affil{Astrophysics Group, Imperial College London, Blackett Laboratory,
               Prince Consort Road, London, SW7 2AZ, U.K.}
\altaffiltext{1}{e-mail: {\tt rubina@ic.ac.uk}.}

\author{Schuyler D. van Dyk } %\altaffilmark{2}}
\affil{Spitzer Science Center, 220-6, Pasadena, CA 91125, USA }

\author{Peter A. H\"{o}flich} %  \altaffilmark{3}}
\affil{McDonald Observatory, University of Texas at Austin, Austin, TX 78712., USA}
\and

\author{Seppo Mattila }%\altaffilmark{4}}
\affil{Stockholm Observatory, Department of Astronomy, AlbaNova, 106 91 Stockholm, Sweden}


%\altaffiltext{1}{Visiting Astronomer, Cerro Tololo Inter-American Observatory.
%CTIO is operated by AURA, Inc.\ under contract to the National Science


\begin{abstract}
We present mid-infrared observations with the {\it Spitzer Space Telescope} of 
the nearby type II-P supernova, SN~2004dj, at epochs of 89 to 129\,days.
We have obtained the first mid-IR spectra of any supernova apart  
from SN~1987A.  A prominent [NiII]\, $\lambda$6.64\,$\mu$m line is
observed, from which we deduce that the mass of stable nickel must be at
least $2.2\times 10^{-4}\,M_\odot$. We also observe the red wing of the
CO-fundamental band. We relate our findings to possible progenitors
and favour an evolved star, most likely a red supergiant, with a probable
initial mass between $\sim$10 and 15\,$M_\odot$.
\end{abstract}


\section{Introduction}

Core-collapse supernovae (SNe) are the end points of most stars more massive
than $\sim$8$M_\odot$. As such, they provide a key test of 
stellar evolution. Furthermore, they play a major
role in driving the chemical and dynamical evolution of galaxies, and
have also been proposed to be major contributors to dust at epochs
when the Universe was still young ($\mathrm{z}\ga6$) \citep[e.g.][]{tf:01}.

SN explosions provide unique natural laboratories for studying,
in real time, the physics of a variety of combustion, hydrodynamic,
nuclear, and atomic processes. While SNe constitute important
astronomical sources over all wavelength ranges, the combination of
strong absorption by the Earth's atmosphere and the high 
background at mid-IR wavelengths
has meant that this region has thus far remained inaccessible for the
study of SNe, apart from the exceptionally nearby SN~1987A
\citep[e.g.][]{roche:93,wooden:93}.

Nevertheless, the mid-IR region holds the potential of providing
unique insights into the nature of SN explosions and the role
played by dust in these events. 
Although abundance measurements have long been carried out using
UV/optical spectra, the large number of lines coupled with strong
Doppler broadening leads to line-blending resulting in ambiguities
in species identification and errors in flux measurement.
In contrast, the fewer line transitions and much-reduced sensitivity
to extinction in the mid-IR, allows firm line identifications
and accurate measurements of intrinsic line strength.
Abundance measurements using fine-structure lines are particularly 
robust as these are largely insensitive to temperature.
Furthermore, warm dust that may condense in the ejecta emits most
strongly in the mid-IR. Moreover, by monitoring the mid-IR spectral
energy distribution and evolution, we may discriminate between pre-existing
circumstellar dust and newly condensing dust in the ejecta.

The advent of the {\it Spitzer Space Telescope\/} \citep[SST;][]{werner:04} 
with its vastly improved mid-IR sensitivity/spatial resolution combination, 
compared with previous instrumentation, has finally opened up the possibility 
of studying typical nearby SNe in the mid-IR.  
In this {\it Letter\/}, we report on the first mid-IR results for SN~2004dj,
the nearest SN in over a decade.

SN~2004dj was discovered in the nearby nearly face-on spiral galaxy
NGC~2403 on 2004 July 31 by K. Itagaki at a visual magnitude of 
+11.2 \citep{nakano:04}. 
It appears within a star cluster \citep[Sandage 96;][]{sandage:84,ma:04}. 
A spectrum obtained on 2004 August 03 by \citet{patat:04} revealed SN~2004dj 
to be a type II-P (plateau) SN at an epoch of $\sim$3 weeks post-explosion.
In what follows, we assume an explosion date of 2004 July 10.
\citet{ko:05} present an $R$-band light curve which shows
that the plateau phase had ended by about +100\,d post-explosion.  NGC~2403 lies
within the M81 group at a distance of 3.13\,Mpc \citep{freedman:01}. 
About a month post-explosion, SN~2004dj was also detected at radio 
and X-ray wavelengths \citep{stockdale:04,pooley:04}.

\begin{figure}[!t]
\begin{centering}
\includegraphics[height=0.45\textwidth,clip=]{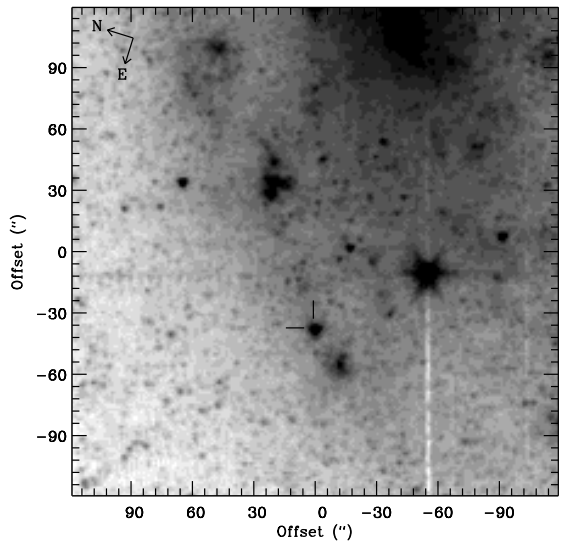}    % referee
\caption{Subsection of a 3.6$\mu$m Spitzer (IRAC) image taken at
         +89\,d showing the immediate field around SN~2004dj. 
         Short dashes mark the position of the SN
         ($\alpha_{\mathrm{J2000}}=07^{h}37^{m}17^{s}.02;\,\,
         \delta_{\mathrm{J2000}}= +65\degr 35'57''.8$) which lies
         10$''$N and 160$''$E of the galaxy nucleus.}
\end{centering}
\label{fig:chart}
\end{figure}

\begin{table*}
\begin{center}
\caption{Photometry of SN~2004dj \label{tab:obslog}}
%\fontsize{10.4}{11}\selectfont
\begin{tabular}{ccclccccccclc}
\tableline\tableline
        &  & & \multicolumn{4}{c}{Flux (mJy)} \\
        &  Epoch & t$_{\mathrm{exp}}$ & \multicolumn{4}{c}{IRAC} &  & Epoch & t$_{\mathrm{exp}}$ & \multicolumn{1}{c}{MIPS} \\
\cline{4-7} \cline{11-11}
Date & (d)      & (s) & \hspace{0.7cm}3.6  & 4.5 & 5.8  & 8.0$\mu$m & Date & (d) & (s) & 24$\mu$m \\
\tableline
\phn2004 Oct. 07         & \phn+89 & 150 & \phn9.99$\pm$0.03$^\ddag$  & 7.60$\pm$0.03 & 6.53$\pm$0.05 & 4.09$\pm$0.04 & $^\dag$2004 Oct. 12 & \phn+94 & \phn41.9 & 1.1$\pm$0.3 \\
$^\dag$2004 Oct. 10  & \phn+92 & 240 & 10.82$\pm$0.03 & 8.64$\pm$0.03 & 6.15$\pm$0.05 & 3.93$\pm$0.04 & \phn2004 Oct. 14 & \phn+96 & 165.7 & 1.2$\pm0.2$ \\
$^\dag$2004 Oct. 12  & \phn+94 & 240 & \phn7.41$\pm$0.03  & 8.43$\pm$0.03 & 5.72$\pm$0.05 & 3.39$\pm$0.04 & $^\dag$2004 Oct. 16 & \phn+98 & \phn41.9 & 1.0$\pm0.3$ \\
\phn2004 Nov. 01         & +114    & 150 & \phn4.56$\pm$0.02  & 6.67$\pm$0.02 & 4.48$\pm$0.04 & 2.48$\pm$0.04 & \phn2004 Nov. 06 & +119    & 165.7 & 1.1$\pm0.2$ \\
\tableline
\end{tabular}
\end{center}
\mytablecomments{The epochs are for an assumed explosion date of 2004
Jul. 10 \citep{patat:04}.  t$_\mathrm{exp}$ is the on-source integration
time. Photometry was carried out using a
3\farcs6 radius aperture for all four IRAC bands.  Aperture
corrections of 1.12, 1.12, 1.14, and 1.23 were applied to channels 1-4
respectively ({\it cf.} Table 5.7 of the IRAC data handbook).  For the
24$\mu$m MIPS data, we used a 5\farcs6 radius aperture
%together with
%two adjacent off-source positions to estimate the background
(aperture correction = 1.72; MIPS data handbook, Fig. 3.2).
Using a 6\farcs0 radius aperture for the MIPS 70$\mu$m channel,
we estimate a rough upper limit of 10\,mJy.  t$_\mathrm{exp}$ at
70 and 160$\mu$m is, respectively, 125.8 and 25.2\,s for our set-up,
and 41.9 and 4.2\,s for the SINGS set-up.
$^\ddag$ The errors shown above are statistical errors only.
Note that systematic errors in the calibration can be as large
at 10\%, although the relative errors are likely to be much smaller.
$^\dag$ SINGS data.  }
\end{table*}

\section{Observations}

SN 2004dj was observed with the SST as part of Director's Discretionary Time
(DDT) Program 226.  Here we report on the first two of four epochs with all
three instruments (IRAC, MIPS, IRS) at epochs of +89 to +139\,d.
Two further epochs of IRAC and MIPS images were serendipitously obtained 
from the Spitzer Investigation of Nearby Galaxies (SINGS) Legacy project 
\citep[Spitzer PID 159;][]{kennicutt:03} in 2004 October.

\subsection{Photometry}

From the IRAC images it is clear that we have detected a source coincident 
with the position of SN~2004dj (Fig. \ref{fig:chart}). That
this source is dominated by emission from the SN is evident
from its fading with time (Table \ref{tab:obslog}). 
Aperture photometry was carried out on the IRAC and MIPS (24$\mu$m)
post-BCD image mosaics using GAIA. We repeated the procedure for two IRAC
epochs (+89 and +114\,d) using the APEX software applied to the BCD
frames and found generally consistent results.  The results of our 
measurements are given in Table \ref{tab:obslog}.  
The SN was not detected in the 70 and 160$\mu$m MIPS channels,
probably due to a combination of high host galaxy background and 
poorer spatial resolution in these channels.

\citet{sugerman:05} have reported the rapid 
(5-day) decline of SN~2004dj at mid-infrared wavelengths.
We point out that between +89\,d and 114\,d the plateau phase
was ending \citep{ko:05} which accounts for the rapid
decline in the 3.6$\mu$m and 8.0$\mu$m channels during this time. The
slower decline at the other two wavelengths is probably due to the
emerging CO fundamental band emission (see \S\,\ref{subsec:co}).
There is also apparent evidence of a slower decline at 24~$\mu$m,
but this only significant at the 80\% level.
We therefore do not discuss it further.

\subsection{Spectroscopy}
Spectra were obtained on 2004 October 24 (+106\,d) and 2004 November 16
(+129\,d) using the Short-Low (SL, $\lambda = 5.2-14.5\mu$m; R =
64-128) and Long-Low (LL, $\lambda = 14-38\mu$m; R = 64-128) modules
of the Infrared Spectrograph. The observations were performed in staring mode. 
Total time-on-source for both
epochs are 609.5\,s in SL and 629.1\,s in LL.  Module slits were positioned
relative to a reference star using a moderate-accuracy peak-up with the blue
(16$\mu$m) channel.

\begin{figure}[!t]
\begin{centering}
\includegraphics[width=0.45\textwidth,clip=]{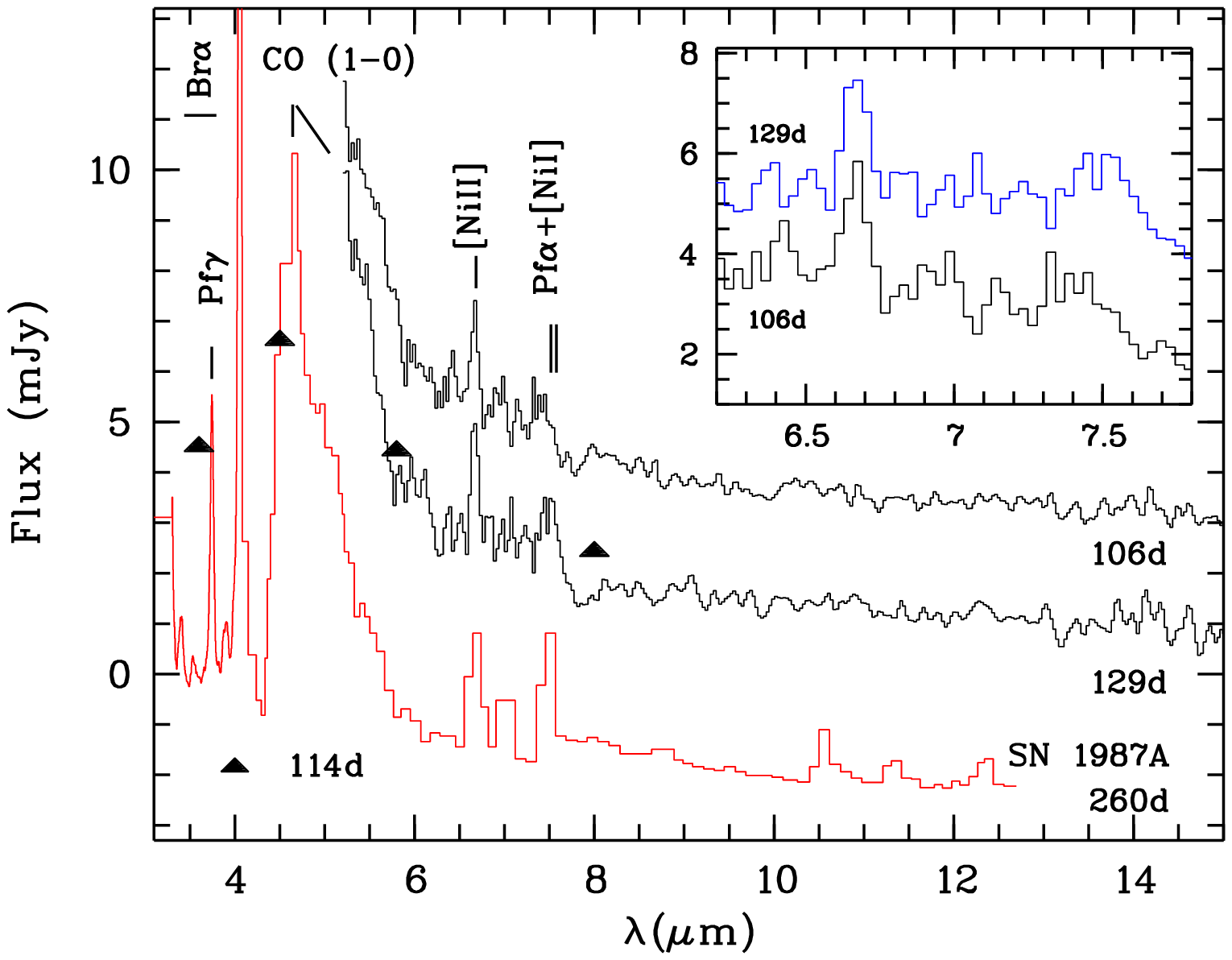}
\caption{IRS spectra of SN~2004dj.  The 106\,d spectrum
  has been shifted by +2.5\,mJy. The lowermost curve is the 
  260\,d spectrum of SN~1987A \citep{meikle:89,wooden:93}; 
  it has been arbitrarily scaled to highlight the striking similarity with SN~2004dj. 
  The 114\,d IRAC photometry is also shown. 
  The inset zooms in on the 6.2--7.8$\mu$m region of the SN~2004dj spectra 
  (the 129\,d spectrum has been shifted by +2.5mJy) showing the profiles of the [NiII] line and 
  the emergence of the blend near 7.5$\mu$m.
  \label{fig:04dj_irs}}
\end{centering}
\end{figure}

The SL data were preprocessed using version S11.0.2 of the 
Spitzer data processing pipeline. 
Subsequent reduction of these post-BCD data was carried out
within the {\sc FIGARO 4\/} environment.  We removed the
background emission by differencing the two nod positions. 
For the first order SL spectrum, the sky position (for both
epochs) unfortunately landed on a cluster of bright sources, rendering it 
unusable. We therefore extracted the spectrum from the single uncontaminated 
nod position. 
We used the `tune' tables for the wavelength calibration.
Flux calibration was carried out in two steps: each order 
was calibrated separately using the FLUXCON
keywords provided. We then merged the spectra, manually clipping the ends
of each order where the noise increases significantly. The spectra were 
scaled to match our day~+114 IRAC channel 3, 4 photometry. 
The IRS epochs are, respectively, 8\,d before and 15\,d after this
IRAC epoch, but no other post-plateau IRAC points were
available. However, the very slow mid-IR evolution of SN~1987A around
these epochs suggests that the fluxing error introduced by the epoch
differences is probably small.  We applied scaling factors of 
of $\times$0.84 and $\times$0.8 to the 106 and 129\,d spectra respectively.  
The uncertainty in these factors is $\sim$15\% which probably dominates 
the overall fluxing error.  We repeated the above reduction sequence 
for the +106\,d spectrum, but this time starting from the BCD data and found 
excellent agreement with the post-BCD results.  The spectra are shown in
Fig. \ref{fig:04dj_irs}.  The SN was not detected in the LL modules.

We supplemented the IRS data with 0.9-2.4\,$\mu$m spectroscopy carried
out on 2004 November 24 (+137d) using the LIRIS instrument mounted at the
Cassegrain focus of the William Herschel Telescope. These data were
obtained using the low-resolution $ZJ$ and $HK$ grisms in the standard
ABBA pattern with 10\arcsec\ nods. The data were reduced in the usual
fashion; the F6 dwarf, BS~3028, was our chosen flux standard.
Final fluxing was achieved by comparison with field stars in the $J$- 
and $H$-band acquisition images, calibrated using 2MASS 
data. The derived SN magnitudes are $J=+13.62\pm0.04$ and
$H=+13.30\pm0.04$.  We used these values to derive scaling factors
($ZJ=\times$1.14 and $HK=\times$2.68) which we applied to the near-IR spectrum
shown in Fig. \ref{fig:nir}.  (The 2MASS pre-explosion magnitudes of the
underlying cluster are $J=+16.04\pm0.11$ and $H=+15.74\pm0.12$).

\section{Analysis}

\subsection{Carbon Monoxide}
\label{subsec:co}
\begin{figure}[!t]
\begin{centering}
\mbox{
\includegraphics[width=0.45\textwidth,clip=]{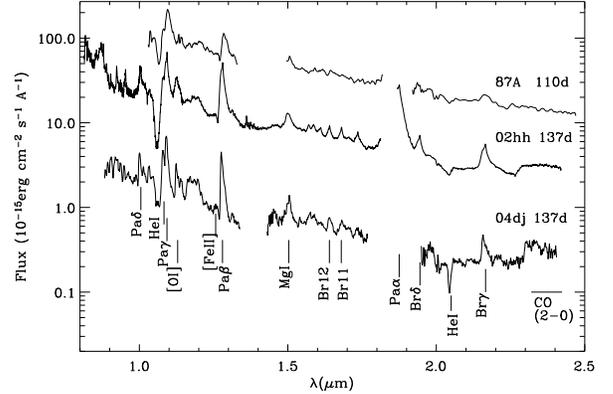}
     }
\caption{Comparison of the near-IR spectra of SN~2004dj with those of
         other, roughly coeval core-collapse SNe.  All spectra
         have been corrected for redshift and reddening.  The
         SN~2002hh spectrum has been taken from \citet{pozzo:05} and
         has been smoothed for display purposes. The SN~1987A spectrum
         has been taken from \citet{meikle:89}.  We used $E(B-V)$ of
         0.18 for SN~2004dj, 2.26 for SN~2002hh, and
         0.19 for SN~1987A. The flux scale is appropriate for
         SN~2004dj only. 
         \label{fig:nir}
         }
\end{centering}
\end{figure}

The two IRS spectra show little difference in overall appearance.
The blue-most region (see Fig. \ref{fig:04dj_irs}), is dominated by a 
rapidly rising slope which we identify with the red wing of the carbon monoxide 
fundamental (4.65\,$\mu$m). This identification is reinforced by the clear 
detection of (i) the first overtone of CO at $\sim2.3\mu$m (Fig. \ref{fig:nir}) 
and (ii) the growth with time of the excess flux in channels 2 and 3 compared
with channels 1 and 4  (see Table \ref{tab:obslog}). 

Molecule formation provides a sensitive diagnostic of the conditions
and degree of mixing in the SN ejecta. Cooling by CO sets the
temperature structure, allowing the temperature to drop to
$\lesssim$1600\,K within a few photospheric radii. As it is optically
thick, the CO-fundamental band forms at several times the photospheric
radius, while the first overtone forms close to the photosphere (defined
to be where the Thomson scattering optical depth = 1). Thus,
the ratio of the fundamental to the first overtone provides a powerful
means for constraining both temperature and density. 
We use our near-simultaneous fundamental and first-overtone 
observations to investigate conditions in the SN~2004dj ejecta. Only a 
summary of first results is given here, as a more detailed analysis will 
be presented elsewhere.

\begin{figure}[!t]
\begin{center}
\includegraphics[width=0.31\textwidth,angle=-90]{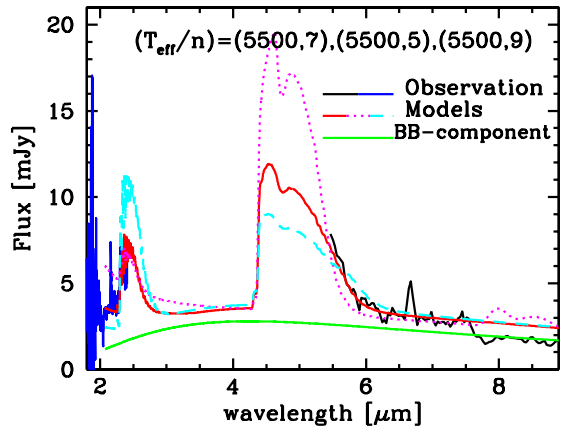}
\caption{Comparison between observed and model profiles for the
vibrational bands of CO on day~130.
The models are identified by the
$(\mathrm{T_{eff}}$,n) tuples (see the colour version of this figure).
To fit the observations, we have added a 1200\,K blackbody component. 
Both $n$ and $\mathrm{T_{eff}}$ have been varied to demonstrate the 
sensitivity of the emitted profiles to the physical conditions. Best agreement 
is obtained with $\mathrm{T_{eff}}=5500\,K$ and $n=7$ (red curve) which are 
close to those of the explosion model (see text).
\label{fig:model}}
\end{center}
\end{figure}

We selected the explosion model from \citet{chieffi:03} that best matched
the plateau duration of SN~2004dj and adjusted the $^{56}$Ni mass to match 
the radioactive tail. The parameters of this model were:
$M=15M_\odot$, explosion energy = $10^{51}$erg, and
$M(^{56}$Ni)=0.03$M_\odot$. Detailed radiative transport calculations
were performed by solving the time-dependent rate equations for the
formation of CO and SiO molecules, following \citet{liu:92}. We recalculated
the temperature structure for the optically thin layers under the assumption 
of radiative equilibrium and evolved the model up to day 130, when the photosphere 
had receded well into the He-enriched layers, 
the structure of which depends on the details of several processes e.g.
mixing. We investigated the sensitivity to various parameters of CO formation, 
by computing an exploratory grid of 36 models with the photospheric temperature, 
photospheric expansion velocity, and density profiles above the recombination zone in the range
$3000 \leqslant \mathrm{T_{eff}(K)} \leqslant 5500$,
$2000 \leqslant \mathrm{v_{exp}(\kms)} \leqslant 4000$, $-3 \leqslant n \leqslant -9$. 
We find a best fit with $\mathrm{T_{eff}}=5500$\,K and $n=7$ (Fig.~\ref{fig:model}),
values which corroborate a 15$M_\odot$ progenitor.

The slope of the spectral energy distribution in the near-IR requires an underlying 
continuum to adequately fit the data (see Fig. \ref{fig:model}). If this is due to dust 
from the SN, then given 
the relative youth of SN~2004dj, it is likely to be due to pre-existing circumstellar dust, 
rather than newly condensing dust. A more likely source of this continuum is 
the underlying cluster. Using Starburst99 models, \citet{ma:04} infer that the cluster 
contains $\approx$12 red supergiants, which is broadly consistent with the near-IR flux from 
our data, as well as the pre-explosion $J$ and $H$ magnitudes.

\subsection{Stable Nickel}

The other conspicuous feature in the SN~2004dj mid-IR spectra is a
prominent emission feature at 6.67$\mu$m which we identify as the
[NiII]~$\lambda 6.634\mu$m fine-structure line. This is produced by the
a$^2$D$_{3/2}$-a$^{2}D_{5/2}$ transition to the ground state. 
That this line is already visible at the start of the radioactive tail,
constrains the degree of mixing of the ejecta.
The line is unresolved, indicating an expansion velocity of $\lesssim 3500\kms$.
Its intensity at 106\,d is
$1.1\pm0.1\times10^{-14}$\,erg\,s$^{-1}$cm$^{-2}$, rising to
$1.5\pm0.15\times10^{-14}$\,erg\,s$^{-1}$cm$^{-2}$ on day~129.  We used
this line to estimate the mass of Ni$^{+}$.  The critical
density for the 6.63\,$\mu$m transition is $1.3\times10^7$cm$^{-3}$
\citep[e.g.][]{wooden:93}. At an epoch of $\sim$100\,d it is likely
that the electron density was above this value.  For example,
\citet{clocchiatti:96} estimate an electron density of $2.5\times
10^{9}$cm$^{-3}$ within 2500\,\kms for the Type~II SN~1992H at
100\,d. On this basis, we deem a simple LTE treatment to be valid. 
A-values were taken from \citet{quinet:96} and partition function values 
from \citet{halenka:01}. 
The Ni$^+$ mass was estimated for 3000\,K and 6000\,K. As expected, the
result was rather insensitive to temperature.  At 106\,d and T=3000\,K we 
obtain M(Ni$^+$)=1.7$\times10^{-4}$~M$_{\odot}$.  Similar values were
obtained with T=6000\,K.  At 129\,d we obtain
M(Ni$^+$)=2.2$\times10^{-4}$~M$_{\odot}$, again with similar values
for both temperatures.  Sobolev optical depths at the two epochs were
$\sim$0.2 and 0.4, respectively. 

Given the clear presence of the [Ni\,II]\,$\lambda$6.634~$\mu$m line,
we examined the spectra for the [Ni\,I]\,$\lambda$7.50\,$\mu$m line
which is produced by the a$^3$F$_3$-a$^3$F$_4$ transition to
ground. While there is little sign of this line at 106\,d, by 129\,d
there is a feature whose red wing corresponds exactly to the
expected location of the 7.50\,$\mu$m line.  We therefore
suggest that the 7.50\,$\mu$m line is present, but blended with another
line to the blue. The most likely candidate is Pf$\alpha$
(7.46$\mu$m).  There may also be contributions to the observed feature
from Hu$\beta$ (7.50$\mu$m) and H$_{7-11}$ (7.51$\mu$m). Consequently
direct measurement of the [Ni\,I]\,$\lambda$7.50\,$\mu$m line
intensity is impractical.  In a future paper, estimates will be made
of the flux contribution from the H~I lines in order to determine the
intensity of the [Ni\,I] line.

%As indicated above, we estimate 0.00022~M$_{\odot}$ for the Ni$^+$
%mass on day~129.  
At 129\,d, virtually all the Ni must be made
up of stable isotopes, dominated by $^{58}$Ni. The derived mass can
be regarded as a firm lower limit for the total stable Ni
mass. As already indicated, it is likely that flux from Ni$^0$ is also
present.  Moreover, the [NiII] line is close to being optically
thick. The presence of a significant continuum suggests
that yet more Ni lies below the photosphere.

%%%%%

\subsection{Constraints on the progenitor mass}

Following \citet{hamuy:03}, we use the $V$-band luminosity of the
exponential tail of the light curve\footnote{ Reported
on http://www.astrosurf.com/snweb2/2004/04dj/04djCurv.htm} 
after 100\,d to estimate the mass of $^{56}$Ni. We find a mean mass 
of $\sim$0.022$\,M_\odot$ which suggests a progenitor mass of $\gtrsim 10\,M_\odot$
\citep{woosley:96} and is consistent with the value used in \S \ref{subsec:co}.

\citet{thielemann:96} predict stable Ni masses for core-collapse SNe having 
progenitor masses of 13--25~M$_{\odot}$. Up to 20\,M$_{\odot}$, they predict 
masses of 0.007--0.013\,M$_{\odot}$, but a much lower value of
0.002\,M$_{\odot}$ for a 25\,M$_{\odot}$ star.  At this early
epoch, our lower limit does not yet constrain this
mass range.  However, as the SN expands, we will be able to 
to obtain a more definitive Ni mass estimate.

By fitting a variety of cluster spectral energy distributions, \citet{ma:04} 
have inferred that Sandage 96 is a 13.6\,Myr cluster with a turn-off mass of
15$M_\odot$. Although they favour a red supergiant progenitor for SN~2004dj, 
they could not completely rule out a blue supergiant progenitor. Our CO analysis
(\S \ref{subsec:co}) supports a red supergiant progenitor.
This is indirectly supported by the lack of
PAH features (at 6.2, 7.7, 8.6, and 11.3\,$\mu$m) which trace the far-UV 
stellar flux and therefore, the young, hot stellar population \citep{peeters:04}.

\vspace{-0.25cm}

\acknowledgments
\noindent
This work is based on observations made with the Spitzer Space Telescope, 
which is operated by the Jet Propulsion Laboratory, California Institute of Technology 
under NASA contract 1407. Support for this work was provided by NASA through an award 
issued by JPL/Caltech. R.K. acknowledges support from the EC Programme `The Physics of Type 
Ia SNe' (HPRN-CT-2002-00303). 
We thank C. Gerardy for useful discussions, and M. Pozzo for providing the SN~2002hh spectrum
(Fig. \ref{fig:nir}).

\end{document}